%% file: pagai_article.tex
\documentclass{article} 

\usepackage{ifxetex}
\ifxetex
\else
\usepackage[utf8]{inputenc}
\fi

\usepackage{amsmath,amsthm}
\usepackage{mathcomp,amsfonts,amssymb}
\usepackage[sort,numbers]{natbib}
\usepackage[mathscr]{euscript}
\usepackage{algorithmic,algorithm}
\usepackage{graphicx}
\usepackage{color}
\usepackage{tikz}

\usepackage{enumitem}
\usepackage{dcolumn}
\usepackage{listings}
\usepackage{dmnatbib}
\usepackage{hyperref}

\hypersetup{pageanchor=false}

\newcommand{\ZZ}{\mathbb{Z}}

\newtheorem{definition}{Definition}

\tikzstyle{arrow}=[->,line width=.05cm,draw=red!90!blue!60!black]

\usetikzlibrary{snakes,arrows,shapes,backgrounds,shadows,automata,patterns}

\tikzstyle{state}=[circle,fill=black!25,minimum size=13pt,inner sep=0pt]
\tikzstyle{rstate}=[rectangle,fill=black!25,minimum size=13pt,inner sep=0pt]
\tikzstyle{transition}=[rectangle,semithick,draw=black!75,
  			  minimum size=4mm]
\tikzstyle{transition2}=[transition,rectangle,thick,dashed,
  			  minimum size=4mm]
\tikzstyle{PRstate}=[circle,double,draw,fill=blue!15,minimum size=13pt,inner sep=0pt]
\tikzstyle{polyhedra}=[blue!25,opacity=0.5,pattern=north west lines,pattern
color=blue]
\tikzstyle{line}=[black,thick]

\lstnewenvironment{C}
{\lstset{language=C}}
{}

\title{PAGAI: a path sensitive static analyzer} 
\author{Julien Henry \and David Monniaux \and Matthieu Moy}

\begin{document}
\maketitle

\begin{abstract} 
	We describe the design and the implementation of PAGAI,
        a new static analyzer working over the LLVM compiler infrastructure,
        which computes inductive invariants on the numerical variables of the
        analyzed program.

	PAGAI implements various state-of-the-art algorithms combining
        abstract interpretation and decision procedures (SMT-solving),
        focusing on distinction of paths inside the control flow graph while
        avoiding systematic exponential enumerations.
        It is parametric in the abstract domain in use, the iteration
        algorithm, and the decision procedure.

	We compared the time and precision of various combinations of
	analysis algorithms and abstract domains, with extensive
        experiments both on personal benchmarks and widely available
        GNU programs.
\end{abstract}

\section{Introduction}\label{intro}
Sound static analysis automatically computes properties on programs, such as the possible values of their variables during execution.
Applications include:
showing that a program cannot encounter a runtime error (such as arithmetic overflow, division by zero, array access out of bounds), as in e.g. the Astrée analyzer \cite{ASTREE_ESOP05};
computing invariants for use with assisted proof systems (such as the B method~\cite{Abrial_B_2005}), thereby lessening the burden on the user;
computing invariants for advanced optimization techniques in compilation (e.g. showing that two array cells are distinct, in order to allow instruction reordering between assignments to these cells).
All these applications need invariants on numerical quantities.

This article introduces PAGAI, a new tool for fully automatic static analysis.
PAGAI takes as input a program in the ``bitcode'' intermediate representation of LLVM \citep{LLVM_langref,Lattner:2004:LCF:977395.977673}, a modern compilation framework.
LLVM bitcode is a target for several industrial-strengh compilers, most notably Clang (supporting C, C++, Objective-C and Objective-C++) and llvm-gcc (supporting, in addition to these, Fortran and Ada);
furthermore, a growing number of analysis tools, testing tools, etc. are currently built around this platform (Calysto, KLEE, LAV, LLBMC).

The output of PAGAI is a list of inductive invariants for a selected subset of the control nodes of the original program:
for structured source programs, PAGAI will provide an inductive invariant for loop headers.%
\footnote{A preliminary analysis pass selects a subset of nodes that cuts all cycles in the control-flow graph, by selecting all targets of return edges in a depth-first search traversal; when applied to a structured program, it selects loop headers.}

At present, PAGAI checks user-specified safety properties provided through assertions using the standard C/C++ \lstinline|assert(condition)| macro. The tool will attempt proving that the assertion failure is unreachable and, if unsuccessful, provide a warning message (the tool does not at present include bounded model checking or path exploration techniques for reconstructing an actual failure trace, thus such a warning message should be interpreted as a \emph{possible} assertion failure). It also allows user-specified assumptions, through the \lstinline|assume(condition)| macro. Executing traces falsifying assertions or assumptions are considered to terminate when executing the macro; thus, user-specified assertions may be used to guide the analyzer by providing invariants that it was not able to synthesize by itself.
Possible extensions could include checking for memory safety of array accesses.

PAGAI is based on abstract interpretation, a general framework for fully automatic static analysis.
PAGAI infers invariants of a selected form; by default it performs \emph{linear relation analysis}, which obtains invariants as conjunctions of linear inequalities (or, equivalently, convex polyhedra), but it also supports other abstract domains through a runtime option.
Depending on the \emph{iteration algorithm} selected, PAGAI may also infer invariants as \emph{disjunctions} of elements of the abstract domain (e.g. unions of convex polyhedra).

Textbook descriptions of abstract interpretation-based static analysis state that an inductive invariant is computed at every control point of the program.
In contrast, PAGAI abstracts straight-line sequences of statements en bloc, computing invariants only at points where control flow branches or merges.
Furthermore, several algorithms implemented in PAGAI compute invariants only at the heads of loops (or, in general control-flow graph, at nodes forming a \emph{feedback vertex set}, whose removal breaks all cycles in the graph), expanding the rest of the control flow to a possibly exponential number of straight-line sequences of statements between the selected nodes.
In order to avoid explicit enumerations of exponential sets, PAGAI uses decision procedures for arithmetic theories, based on the \emph{satisfiability modulo theory} (SMT) approach: each path is enumerated only if needed, in response to a positive satisfiability query~\cite{Monniaux_Gonnord_SAS11,Henry_Monniaux_Moy_SAS2012}.

The PAGAI tool is dedicated to experimenting with new analysis algorithms.
It allows independent selection of the abstract domain and the iteration strategy, and partially independent selection of decision procedure,%
\footnote{Certain abstract domains express relationships, such as linear congruences, that certain decision procedures cannot deal with. It is at present necessary that the decision procedure reflects semantics at least as precise as those of the abstract domain. This limitation will be lifted in the future.}
and thus is well-suited for comparisons.
We thus conducted extensive experiments both on examples we produced ourselves (sometimes inspired by industrial code) and on GNU programs,
for which the ability to run on any C or C++ code, through the LLVM system, was especially useful.
Front-ends for many analysis tools put restrictions (e.g. no backward \lstinline|goto| instructions, no pointer arithmetic...), often satisfied by safety-critical embedded programs, but not by generic programs;
our tool suffers no such restrictions, though it may in some cases apply coarse abstractions which may possibly yield weak invariants.

After illustrating the limitation of traditional abstract
interpretation on an example in section~\ref{sec=motivating-example},
we will describe PAGAI's implementation in
section~\ref{sec=implementation}, and comment on the results of
extensive experiments in section~\ref{sec:experiments}, allowing the
comparison of state of the art techniques on real-life programs.

\section{Motivating Example}
\label{sec=motivating-example}
In most forward abstract interpretation-based analyses, when control flows from several nodes into a single node, the abstract value at that node is obtained by computing the least upper bound of the incoming abstract values in the abstract domain (in backward analysis, this occurs when control flows from a single node to several nodes).
If the abstract domain is convex polyhedra, then this means computing the convex hull of the incoming polyhedra.
Such an operation may induce unrecoverable loss of precision by introducing spurious states that cannot occur in concrete program runs.

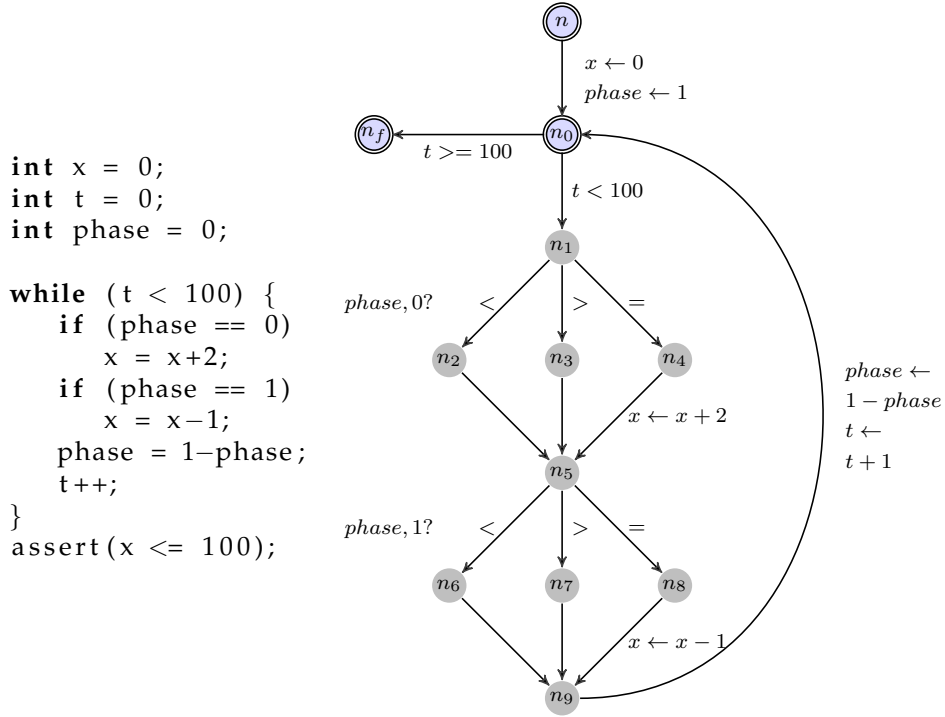
\begin{figure}[!h]
\begin{minipage}[c]{.35\linewidth}
\begin{C}
int x = 0;
int t = 0;
int phase = 0;

while (t < 100) {
   if (phase == 0) 
      x = x+2;
   if (phase == 1) 
      x = x-1;
   phase = 1-phase;
   t++;
}
assert(x <= 100);
\end{C}
\end{minipage}
\begin{minipage}[c]{.64\linewidth}
\begin{tikzpicture}[->,>=stealth',auto,node distance=1.5cm,
                    semithick,font=\footnotesize]

	\node[PRstate] (n0) {$n$};
	\node[PRstate] (n00) [below of=n0] {$n_0$};
	\node[PRstate] (nf) [left of=n00, node distance=2.5cm] {$n_f$};
	\node[state] (n1) [below of=n00] {$n_1$};
	\node[state] (n3) [below of=n1] {$n_3$};
	\node[state] (n2) [left of=n3] {$n_2$};
	\node[state] (n4) [right of=n3] {$n_4$};
	\node[state] (n5) [below of=n3] {$n_5$};
	\node[state] (n7) [below of=n5] {$n_7$};
	\node[state] (n6) [left of=n7] {$n_6$};
	\node[state] (n8) [right of=n7] {$n_8$};
	\node[state] (n9) [below of=n7] {$n_9$};

  \path [transition] 
		(n0) edge              node  {$\begin{array}{l}
		x \leftarrow 0 \\
		phase \leftarrow 1 
		\end{array}$} (n00);
  \path [transition] 
        (n00)  edge              node {$t>=100$} (nf);
  \path [transition] 
        (n00)  edge              node {$t<100$} (n1);
  \path [transition] 
        (n1) edge			   node [left] {$phase,0?$~ ~ ~ $<$} (n2);
  \path [transition] 
        (n1)  edge              node {$>$} (n3);
  \path [transition] 
        (n1)  edge              node [right] {$=$} (n4);
  \path [transition] 
        (n3) edge              node  {} (n5);
  \path [transition] 
        (n2) edge			   node {} (n5);
  \path [transition] 
        (n4) edge			   node [right] {$x \leftarrow x+2$} (n5);
  \path [transition] 
        (n5) edge			   node [left] {$phase,1?$ ~ ~ ~$<$} (n6);
  \path [transition] 
        (n5) edge			   node {$>$} (n7);
  \path [transition] 
        (n5) edge			   node [right] {$=$} (n8);
  \path [transition] 
        (n6) edge              node {} (n9);
  \path [transition] 
        (n7) edge              node {} (n9);
  \path [transition] 
        (n8) edge              node [right] {$x \leftarrow x-1$} (n9);
  \path [transition] 
        (n9) edge [out=0, in=0, distance=4.3cm] node [right] {
		$\begin{array}{l}
			phase \leftarrow \\
			1-phase \\
			t \leftarrow \\
			t+1
		\end{array}$} (n00);

\end{tikzpicture}
\end{minipage}
\caption{Example of program, where the loop behaviour vary depending on a variable $phase$.}
\label{example}
\end{figure}

An example of program where such a loss of precision occurs is depicted in
Fig.~\ref{example}.
In this program, the loop body has two feasible paths that
are executed alternatively, depending on a variable ``phase''.
Such programs, with active code paths depending on global ``mode'' or ``phase'' variables, often occur in reactive systems.

Removing program point $n_0$ breaks all cycles; we are thus primarily concerned with obtaining an inductive invariant at that point.
We consider the domain of convex polyhedra and thus wish to obtain this invariant as a polyhedron.
Because convex polyhedra form a lattice of infinite height, we use Kleene iterations (pushing abstract values through control-flow edges) with a widening scheme, which ensures convergence in finite time to an inductive invariant, followed by decreasing (narrowing) iterations.

At program point $n_5$, classical forward abstract interpretation with convex polyhedra computes the convex hull of three incoming polyhedra over variables $(\lstinline|phase|,\lstinline|x|,\lstinline|t|)$. This convex hull introduces extra states, unreachable in the concrete programs, for the analysis of the fragment from $n_5$ to~$n_9$. When analyzing the whole loop, these extra states prevent proving $x < 100$.

To cope with this problem, a solution is to compute disjunctive invariants at all intermediate nodes: at $n_5$, keep an explicit list of three polyhedra, and thus obtain a list of nine polyhedra at $n_9$. We pass the convex hull of these polyhedra to the widening operator at point $n_0$ (which operates on polyhedra, not on lists of polyhedra). The drawback is that the number of elements in the lists may grow exponentially with the number of successive tests.

A second solution, equivalent to the preceding with respect to final results but different in its operation, is to distinguish all nine paths inside the loop (from $n_0$ to $n_0$), compute the final outcome of each path, and compute the convex hull of these outcomes.
Instead of enumerating all nine paths explicitly, we consider them in succession, only as needed. We start with an empty polyhedron at $n_9$ (more generally, it should contain initial states at this control point), and process paths as long as they make this polyhedron grow.
The next path to consider is obtained from a model of an arithmetic formula expressing this growth condition \cite{Monniaux_Gonnord_SAS11}; if this formula is unsatisfiable, this means there is no such path and thus the convex hull encompasses the outcome of all paths.

The advantages of this second method over the preceding one are twofold: there is no exponentially large list of abstract elements, and the satisfiability query for the formula is handed over to a \emph{satisfiability modulo theory} (SMT) solver.
Modern SMT-solvers are very efficient; their caching mechanisms may, for instance, remember that taking a certain branch in the code is incompatible with taking another one (if a Boolean is associated with passing through each branch, then this is just a \emph{blocking clause} inside the underlying SAT-solver).
The algorithms implemented in PAGAI are variants of this approach of implicitly representing of exponentially-sized sets of paths and enumerating them as needed.

\section{Implementation}
\label{sec=implementation}

PAGAI is a prototype interprocedural static analyzer, that implements our
recent combined techniques \cite{Henry_Monniaux_Moy_SAS2012}
as well as the classical abstract interpretation algorithm, and the state-of-the-art techniques \emph{Path Focusing} \cite{Monniaux_Gonnord_SAS11} and \emph{Guided Static Analysis} \cite{DBLP:conf/sas/GopanR07}.

Abstract domains are provided by the APRON library
\citep{DBLP:conf/cav/JeannetM09}, and include convex polyhedra (from the builtin
Polka ``PK'' library), octagons, and products of intervals. It also has
an interface with the Parma Polyhedra Library \cite{BagnaraHZ08SCP}, giving access to
more abstract domains (e.g. a reduced product of polyhedra and linear congruences, producing invariants such as $0 \leq x \leq 1001 \land x \equiv 0 \pmod 7$).

For SMT-solving, our analyzer uses
Yices \cite{DBLP:conf/cav/DutertreM06}
or Microsoft Z3 \cite{DBLP:conf/tacas/MouraB08}
through their C API.
An implementation of communications with the SMT-solver by textual messages sent through a pipe following the SMT-Lib~2 standard \cite{BarST-SMT-10} is underway, and now partially supports Z3, MathSAT~5 and SMTinterpol.%
\footnote{%
It is unfortunately impossible to ignore differences between solvers behind the supposedly standard interface, since different solvers may support slightly different sets of operators and settings and may return models in different formats.}

\begin{table}
\begin{center}\small
\begin{tabular}{|l|r|r|} \hline
	\multicolumn{1}{|c|}{Name} &
        \multicolumn{1}{c|}{kLOC} &
        \multicolumn{1}{c|}{$|P_R|$} \\ \hline
	a2ps & 55 & 2012\\
	gawk & 59 & 902\\ 
	gnuchess & 38 & 1222\\ 
	gnugo & 83 & 2801\\
	grep & 35 & 820\\
	gzip & 27 & 494\\
	lapack/blas & 954 & 16422\\
	make & 34 & 993\\ 
	tar & 73 & 1712\\
	\hline
\end{tabular}
\end{center}

\caption{List of analyzed open-source projects, with their respective number of
lines of code, and their number of control points in $P_R$}
\label{tab:projects}
\end{table}

\subsection{Analysis algorithm}
\label{sec:analysis-algorithm}
For each program, we distinguish a set $P_W$ of suitable widening points by a simple algorithm: initialize $P_W=\emptyset$ and for each procedure, compute the strongly connected components of its control-flow graph using Tarjan's algorithm; the targets of the back-edges of the depth-first search are added to~$P_W$.
The resulting \emph{cut set} or \emph{feedback vertex set} is not necessarily minimal, but is sufficient to disconnect all cycles --- more sophisticated techniques are discussed in e.g. \citet{BourdonclePhd}.%
\footnote{
It would be possible to obtain a feedback vertex set minimal with respect to inclusion by successive removal of nodes. Obtaining one of minimal cardinality is an NP-complete problem, but \citet{DBLP:journals/siamcomp/Shamir79} showed that it can be done in linear time for a class of graphs including reducible graphs, that is, those obtained from structured programs. This latter algorithm is being implemented.}
It is however unclear whether more advanced selection techniques would finally yield stronger invariants; the current simple scheme has the advantage that, when run over a control-flow graph obtained from a structured program, it marks heads of loops, which is a ``natural'' choice.

While \cite{Henry_Monniaux_Moy_SAS2012,Monniaux_Gonnord_SAS11} provide for another set $P_R \supseteq P_W$, with abstract join operators (as opposed to widenings) being applied at points in $P_R \setminus P_W$, our tool does not currently such technique, which is meant to reduce the complexity of SMT formulas at the expense of analysis precision.

LLVM bitcode is in \emph{static single assignment} (SSA) form: a given scalar variable is given a value at a single syntactic point in the program. In concrete terms, an assignment \lstinline|x=2*x+1;| gets translated into a definition $x_2 = 2x_1+1$, with distinct variables $x_1$ and $x_2$ corresponding to the same original variable \lstinline|x| at different points in the program.
Because LLVM generally assigns rather straightforward names (e.g. \lstinline|x.0| for the first renaming of variable \lstinline|x|), the user can map the invariants back to the original source code; an automatic and more robust back-to-source mapping, based on debugging information, is being developed.

LLVM makes it easy to follow definition-use and use-definition chains: for a given variable (say, $x_2$) one can immediately obtain its definition (say, $2x_1+1$).
One may see conversion to SSA form as a static precomputation of some of the symbolic propagations proposed by \citet{DBLP:conf/vmcai/Mine06} to enhance the precision of analyses.

SSA introduces $\phi$-functions at the head of a control code to define variables whose value depends on which incoming edge was last taken to reach this control node. For instance, for \lstinline|if (...) { x = 2*x+1; } else { x= 0; }|, then $x_2$ is defined as $\phi(2x_1+1,0)$.

In this framework, each variable is uniquely defined as an arithmetic ($+$, $-$, $\times$, $/$) function of other variables that themselves may not be representable as arithmetic functions, because they are defined using $\phi$-functions, loads from memory, return values from function calls, or other numerical operations (e.g. bitwise operators) that are not representable with our class of basic arithmetic operations. We may vary the class of arithmetic operations, for instance, by restricting ourselves to linear ones.

This motivates a key implementation decision of our tool: only those variables
$v_1,\dots,v_n$ that are not defined by arithmetic operations are retained as
coordinates in the abstract domain (e.g. as dimensions in polyhedra), assuming
they are live at the associated control point. 

	For instance, assume that $x,y,z$ are numerical variables of a program,
	$x$ is defined as $x = y+z$, and $x,y,z$ are live at point $p$. Instead of having
	$x$ as a dimension for the abstract value at point $p$, we only have $y$ and $z$. All the properties
	for $x$ can be directly extracted from the abstract value attached to $p$ and the relation $x=y+z$.
	This is an optimisation in the sense that there is redundant information in
	the abstract value if both $x,y$ and $z$ are dimensions of $X_p$.
	The classical definition of liveness can be adapted to our case:

	\begin{definition}[Liveness by linearity]
	A variable $v$ is \emph{live by linearity} at a control point $p$ if and
	only if one of these conditions holds:
		(i) $v$ is live in $p$.
        (ii) There is a variable $v'$, defined as a linear combination of other
		variables $v_1, v_2, \dots, v_n$, so that $\exists i \in \{1,\dots,n\}, v = v_i$,
		and $v'$ is live by linearity in $p$.
	\end{definition}

	Finally, a variable is a dimension in the abstract domain if and only if it
	is live by linearity and it is not defined as a linear combination of
	program variables.

A basic block of code therefore amounts to a \emph{parallel assignment} operation between live-by-linearity variables
$(v_1,\dots,v_n) \allowbreak\mapsto\allowbreak
(f_1(v_1,\dots,v_n), \allowbreak, \dots, \allowbreak
 f_n(v_1,\dots,v_n))$;
such operations are directly supported by APRON. This has three benefits:
(i) it limits the number of dimensions in the abstract values, since polyhedra libraries typically perform worse with higher dimensions;%
\footnote{The additional dimensions express linear equalities between variables, which are directly handled by polyhedra library.
They should therefore cost little assuming some sparse representation of the constraints.
Alas, several libraries, including APRON, compute with \emph{dense} vectors and matrices, which means that any increase in dimensions slows computations.}
(ii) the abstract operation for a single path in path-focusing methods also is a (large) parallel assignment;
(iii) as suggested by \citet{DBLP:conf/vmcai/Mine06}, this approach is more precise than running abstract operations for each program line separately:
for instance, for \lstinline|y=x; z=x-y;| with precondition $x \in [0,1]$, a line-by-line interval analysis obtains $y \in [0,1]$ and $z \in [-1,1]$ while our ``en bloc'' analysis symbolically simplifies $z = x - x = 0$ and thus $z \in [0,0]$.

In the event that a node is reachable only by a single control-flow edge (which may occur because of dead code, or during the first phases of guided static analysis), the $\phi$ operation reduces to a copy of the values flowing from that edge. In this case, our tool just propagates symbolic values through the predecessor node, without introducing $\phi$-variables.

\subsection{Use}

PAGAI takes as input an LLVM bitcode file, and outputs an inductive invariant
for each control point in $P_R$ (typically, the widening points).
When a program contains an \lstinline|assert(...)| function call, PAGAI also outputs whether
the statement has been proved.
It is also possible to add some preconditions about the variables, etc, using a
function \lstinline|assume(...)|.
Both \lstinline|assert| and \lstinline|assume| are implemented as C macros.
\lstinline|assert(x)| is roughly defined as \lstinline|if (! x) __assert_fail();|, and the tool just tests for the reachability of \lstinline|__assert_fail();|: if it is unreachable, then the assertion is true.
\lstinline|assume| works with the same principle, and is defined as \lstinline|if (! x) __assumption_declared()|. Both \lstinline|__assert_fail| and
\lstinline|__assumption_declared| are \emph{noreturn} functions, terminating the program immediately.

\subsection{Current limitations of the tool, possible future works}

Our tool currently only operates over scalar variables from the SSA representation and thus cannot directly cope with arrays or memory accessed through pointers. We therefore run it after the ``memory to registers'' (\texttt{mem2reg}) optimization phase in LLVM, which lifts most memory accesses to scalar variables.
The remaining memory reads are treated as nondeterministic choices, and writes are ignored. This is a sound abstraction.%
\footnote{As rightly pointed out by a referee, this is a sound abstraction only if memory safety is assumed. The \texttt{mem2reg} preprocessing phase also assumes memory safety, as well as, possibly, the absence of other undefined behaviors as defined by the C standard. This is the price of using the front-end from a generic compiler: C compilers have the right to assume that undefined behaviors do not occur, including in preprocessing and optimization phases.}

The analysis is currently intraprocedural: function calls are ignored in a sound way (the return value is a nondeterministic choice, the value of all variables escaping from the local scope is discarded...).
In order to increase precision, we apply function inlining as an LLVM optimization phase.
Plans for interprocedural analysis include computing input/output summaries for functions as elements of the abstract domain (e.g. if the function operates over variables $x$ and $y$, then one could compute a polyhedron over $(x,y,x',y')$ encompassing all input-output pairs) or as more general formulas.

Since it is often advantageous to distinguish whether a loop has been executed at least once,%
\footnote{Consider the very simple loop \lstinline|for(int i=0; i<n; i++) {}|.
The obvious loop invariant is $0 \leq \lstinline|i| \leq \lstinline|n|$, but it is valid only if $\lstinline|n| > 0$.
One would thus need to use disjunctive loop invariants to obtain
$0 \leq \lstinline|i| \leq \lstinline|n| \lor
 (\lstinline|i|=0 \land \lstinline|n| \leq 0)$.
It is much simpler to unroll the loop once.}
we unroll every loop once, again with a LLVM optimization phase.

Our tool currently assumes that integer variables are unbounded mathematical integers ($\ZZ$) and floating-point variables are real (or rational) numbers. Techniques for sound analysis of bounded integers, including with wraparound, and of floating-point operations have been developed in e.g. the Astr\'ee system \citep{ASTREE_ESOP05,ASTREE_PLDI03}, but porting these techniques to our iteration schemes using SMT-solving requires supplemental work.
It is unclear whether one should use bitvector arithmetic inside the SMT formula, or use mathematical integers with explicit splits for wraparound.%
\footnote{E.g. an operation $z = x+y$ over $n$-bit signed integers would appear as the disjunction of three statements $z = x+y \land -2^{n-1} \leq x+y < 2^{n-1}$,
$z = x+y+2^n \land -x+y < -2^{n-1}$,
$z = x+y-2^n \land x+y \geq 2^{n-1}$:
one ``normal'' control path and two ``overflow'' paths.}

Our implementation of path-focusing currently does not use true acceleration
techniques, as proposed by \citet{Monniaux_Gonnord_SAS11}. Instead, it simply runs widening and narrowing iterations on a single path.

We currently analyze each strongly connected component of the control-flow graph in topological order; thus each loop nest gets analyzed as a single fixed point.
An alternative method would be to recursively decompose the strongly connected components (for structured programs, this amounts to reconstructing the nested loop structure) and summarize the inner loops before analyzing the outer loop.

The analysis is currently only forward, even though nothing in the techniques
implemented is specific to forward analysis. A possible extension would therefore be backward analysis from the \lstinline|__assert_fail()| statements.

\section{Experiments}
\label{sec:experiments}

We conducted extensive experiments on real-life programs in order to compare the
different techniques, mostly on open-source projects (Tab.~\ref{tab:projects}) written in C, C++ and Fortran.

\subsection{Precision of the various techniques}


For each program and each pair $(T_1,T_2)$ of analysis techniques, we list
the proportion of control points in $P_R$ where $T_1$ (resp.~$T_2$) gives a
strictly stronger invariant, denoted by $\subsetneq$ (resp. $\supsetneq$),
and the proportion of control points where the invariants given by $T_1$ and
$T_2$ are uncomparable for the inclusion ordering (the remainder of the
control points are thus those for which both techniques give the same
invariant). We use convex polyhedra as the abstract domain.

Let us briefly comment the results given in more details in Table~\ref{tab:techniques}.
\emph{Guided Static Analysis} from \citet{DBLP:conf/sas/GopanR07} improves the
result of the classical Abstract Interpretation in $2.21\%$ of the control points
in $P_R$.
\emph{Path-focusing} from \citet{Monniaux_Gonnord_SAS11} 
finds better invariants in $4.13\%$ of the control points.

However, these two techniques also lose precision in an important number
($4.64\%$ for G, $5.14\%$ for PF) of control points, and obtain worse results
than the classical many times. This result is unexpected, and could be partially
explained by bad behaviour of the widening operator.

\begin{table}
	\tiny
\begin{center}
\setlength{\tabcolsep}{0.75ex}
\begin{tabular}{|l
|D{.}{.}{2}D{.}{.}{2}D{.}{.}{2}%
|D{.}{.}{2}D{.}{.}{2}D{.}{.}{2}%
|D{.}{.}{2}D{.}{.}{2}D{.}{.}{2}%
|D{.}{.}{2}D{.}{.}{2}D{.}{.}{2}%
|D{.}{.}{2}D{.}{.}{2}D{.}{.}{2}%
|D{.}{.}{2}D{.}{.}{2}D{.}{.}{2}|} \hline
\multicolumn{1}{|c|}{\textbf{Benchmark}}
& \multicolumn{3}{c|}{\textbf{G/S}}
& \multicolumn{3}{c|}{\textbf{PF/S}}
& \multicolumn{3}{c|}{\textbf{PF/G}}
& \multicolumn{3}{c|}{\textbf{G+PF/PF}}
& \multicolumn{3}{c|}{\textbf{G+PF/G}}
& \multicolumn{3}{c|}{\textbf{DIS/G+PF}} \\ 
& \multicolumn{1}{c}{$\subsetneq$} & \multicolumn{1}{c}{$\supsetneq$} & \multicolumn{1}{c|}{unc.}
& \multicolumn{1}{c}{$\subsetneq$} & \multicolumn{1}{c}{$\supsetneq$} & \multicolumn{1}{c|}{unc.}
& \multicolumn{1}{c}{$\subsetneq$} & \multicolumn{1}{c}{$\supsetneq$} & \multicolumn{1}{c|}{unc.}
& \multicolumn{1}{c}{$\subsetneq$} & \multicolumn{1}{c}{$\supsetneq$} & \multicolumn{1}{c|}{unc.}
& \multicolumn{1}{c}{$\subsetneq$} & \multicolumn{1}{c}{$\supsetneq$} & \multicolumn{1}{c|}{unc.}
& \multicolumn{1}{c}{$\subsetneq$} & \multicolumn{1}{c}{$\supsetneq$} & \multicolumn{1}{c|}{unc.} \\
 \hline
 \input{table1}
\hline
\end{tabular}
\end{center}
\caption{Results of the comparison of the various techniques described in this
paper: classic Abstract Interpretation (S), \emph{Guided Static Analysis} (G),
\emph{Path-focusing} (PF), our combined technique (G+PF), and its version using
disjunctive invariants (DIS). For
instance, \textbf{G/S} compares the benefits of \emph{Guided Static Analysis}
over the classic Abstract interpretation algorithm.
$\subsetneq$ (resp.~$\supsetneq$) gives the percentage of invariants stronger (more precise; smaller with respect to inclusion) with the left-side (resp. right-side) technique,
and ``uncomparable'' gives the percentage of invariants that are uncomparable, i.e neither greater nor smaller;
the code points where both invariants are equal make up the remaining percentage}
\label{tab:techniques}
\end{table}

Finally, our combined technique gives the most promising results, since it is
statistically more precise than the other techniques. It improves the precision
of the inductive invariant in $8.29\%$ to $9.86\%$ of the control points
compared to the three previous techniques. Still, we obtain worse result in a
non-negligible number of cases ($2.02\%$).

The analysis using disjunctive invariants greatly improves the
precision of the analysis (for $14.46\%$ of the control points in $P_R$ compared
to G+PF), at the expense of a higher time cost (see Table \ref{tab:time}).
It also gives worse results in $6.85\%$ of the points, most probably because of
a non-optimal choice of the $\sigma$ function, detailed in
\cite{Henry_Monniaux_Moy_SAS2012}.

While experimenting with techniques that use SMT-solving, we encountered some
limitations due to non-linear arithmetic in the analyzed programs. Indeed, 
the SMT-solver is not able to decide the satisfiability of some SMT-formulae
expressing the semantics of non-linear programs. 
In this case, we skipped the functions for which the SMT-solver returned the
``unknown'' result.
This limitation occurred very rarely in our experiments, except for the analysis
of \emph{Lapack/Blas}, where 798 over the 1602 functions have been skipped.
\emph{Lapack/Blas} implements matrix computations, which use floating-point multiplications.
In cases where the formula is expressed in too rich a logic for the SMT-solver to deal with, a number of workarounds are possible, including:
(i) \emph{Linearization}, as per \citet{DBLP:conf/vmcai/Mine06}, which overapproximates nonlinear semantics by linear semantics.
(ii) Replacing the results of nonlinear operations by ``unknown''.
Neither is currently implemented in our tool.

Table \ref{tab:time} gives the execution time of the different analysis
techniques. It is interesting to see that
\emph{Path-focusing} is sometimes faster than the classical
algorithm.
This seems due to the fact that this algorithm
computes inductive invariant on a small number of control points compared to
classical approaches, thus leading to fewer operations over abstract values.

\begin{table}[!tb]
\begin{center}\small
\begin{tabular}{|l|r|r|r|r|r|} \hline
	\multicolumn{1}{|c|}{Benchmark} &
        \multicolumn{1}{c|}{\textbf{S}} &
        \multicolumn{1}{c|}{\textbf{G}} &
        \multicolumn{1}{c|}{\textbf{PF}} &
        \multicolumn{1}{c|}{\textbf{G+PF}} &
	\multicolumn{1}{c|}{\textbf{DIS}} \\ \hline
	\input{table_time.tex} \hline
\end{tabular}
\end{center}
\caption{Execution time for each technique, expressed in seconds}
\label{tab:time}
\end{table}

\subsection{Precision of Abstract Domains}
\label{sec:compare_domains}
\begin{table}[!tb]
	\tiny
\begin{center}
\setlength{\tabcolsep}{0.75ex}
\begin{tabular}{|l
|D{.}{.}{2}D{.}{.}{2}D{.}{.}{2}%
|D{.}{.}{2}D{.}{.}{2}D{.}{.}{2}%
|D{.}{.}{2}D{.}{.}{2}D{.}{.}{2}%
|D{.}{.}{2}D{.}{.}{2}D{.}{.}{2}%
|D{.}{.}{2}D{.}{.}{2}D{.}{.}{2}%
|D{.}{.}{2}D{.}{.}{2}D{.}{.}{2}|} \hline
\multicolumn{1}{|c|}{\textbf{Benchmark}}
& \multicolumn{3}{c|}{\textbf{PK/OCT}}
& \multicolumn{3}{c|}{\textbf{PK/BOX}}
& \multicolumn{3}{c|}{\textbf{OCT/BOX}}
& \multicolumn{3}{c|}{\textbf{PK/PKEQ}}
& \multicolumn{3}{c|}{\textbf{PK/PKGRID}}
& \multicolumn{3}{c|}{\textbf{POLY/POLY*}} \\ 
& \multicolumn{1}{c}{$\subsetneq$} & \multicolumn{1}{c}{$\supsetneq$} & \multicolumn{1}{c|}{unc.}
& \multicolumn{1}{c}{$\subsetneq$} & \multicolumn{1}{c}{$\supsetneq$} & \multicolumn{1}{c|}{unc.}
& \multicolumn{1}{c}{$\subsetneq$} & \multicolumn{1}{c}{$\supsetneq$} & \multicolumn{1}{c|}{unc.}
& \multicolumn{1}{c}{$\subsetneq$} & \multicolumn{1}{c}{$\supsetneq$} & \multicolumn{1}{c|}{unc.}
& \multicolumn{1}{c}{$\subsetneq$} & \multicolumn{1}{c}{$\supsetneq$} & \multicolumn{1}{c|}{unc.}
& \multicolumn{1}{c}{$\subsetneq$} & \multicolumn{1}{c}{$\supsetneq$} & \multicolumn{1}{c|}{unc.} \\
 \hline
 \input{table_domain.tex}

\end{tabular}
\end{center}
\caption{Results of the comparison of the various abstract domains, when using
the same technique (G+PF). We used as abstract domains Convex Polyhedra (PK and
POLY),
Octagons (OCT), intervals (BOX), linear equalities (PKEQ) and the reduced
product of NewPolka convex polyhedra with linear congruences
from the Parma Polyhedra Library \cite{BagnaraHZ08SCP}.
(PKGRID). The last column compares the domain of Convex Polyhedra with the
improved widening operator from \citet{BagnaraHRZ05SCP} (POLY*), and Convex Polyhedra
using the classical widening operator (POLY). POLY and POLY* use the PPL\cite{BagnaraHZ08SCP}.
$\subsetneq$, $\supsetneq$ and ``unc.'' are defined as in Tab.~\ref{tab:techniques}.}
\label{tab:domain}
\end{table}

For each program and each pair $(D_1,D_2)$ of abstract domains, we compare by
inclusion the invariants of the different control points in $P_R=P_W$ (Tab.~\ref{tab:domain}).

Statistically, the domain of convex polyhedra gives the better results, but
commonly yields weaker invariants than the domains of octagons/intervals;
this is a known weakness of its widening operator~\cite{Monniaux_LeGuen2011}. 
The Octagon domain appears to be much better than intervals; this is
unsurprising since in most programs and libraries, bounds on loop indices are
non constant: they depend on some parameters (array sizes etc.).

The Lapack/Blas benchmarks are unusual compared to the other programs. These
libraries perform matrix computations, using nested loops over indices; such
programs are the prime target for polyhedral loop optimization techniques and it
is therefore unsurprising that polyhedra and octagons perform very well over them.

The analysis of linear equalities (PKEQ) performs very fast compared
to other abstract domains, but yields very imprecise invariants:
it only detects relations of the form $\sum_i a_i x_i = C$ where $a_i$ and $C$ are constants.

Using the reduced product of convex prolyhedra with linear congruences
(PKGRID) improves the analysis by $2.52\%$. 

Finally, we evaluated the benefits of the improved version of the widening
operator for convex polyhedra from \citet{BagnaraHRZ05SCP}, compared to the
classical widening. We found that the improved version from
\citet{BagnaraHRZ05SCP} yields more precise invariants for $3.70\%$ of the
control points in $P_R$.

\subsection{Future Work}
It is not totally relevant to compare by inclusion the abstract values obtained
by the various analysis techniques. Indeed, a slightly smaller invariant may not
always be useful to prove the desired properties.
Future work should thus include experiments with
better comparison metrics, such as
(i) the number of \emph{assert} that have been proved in the code.
		Unfortunately, it is difficult to find good benchmarks or real life
		programs with many \emph{assert} statements;
(ii) the number of false alarms in a client analysis that detects array
		bound violations, arithmetic overflows, etc.

\phantomsection\addcontentsline{toc}{section}{References} 
{
\bibliographystyle{dmabbrvnat}
\bibliography{pagai}
}
\end{document}

%% file: table1.tex
a2ps-4.14      & 0.28 & 0    & 0    & 4.82  & 2.55  & 2.27 & 4.54  & 2.55  & 2.27 & 6.81  & 0.28 & 0    & 8.23  & 0    & 0    & 13.06 & 3.40  & 0.56 \\ 
gawk-4.0.0     & 4.62 & 0    & 0    & 3.70  & 20.37 & 0.92 & 0.92  & 22.22 & 0    & 22.22 & 0    & 0    & 11.11 & 2.77 & 0    & 16.66 & 2.77  & 0.92 \\ 
gnuchess-6.0.0 & 1.51 & 3.47 & 0    & 6.50  & 4.33  & 0    & 6.72  & 3.25  & 0.21 & 6.72  & 2.38 & 0    & 10.19 & 2.38 & 0    & 15.18 & 2.81  & 3.03 \\
gnugo-3.8      & 0.51 & 4.44 & 0.34 & 11.45 & 4.27  & 3.07 & 12.13 & 4.27  & 2.73 & 10.25 & 3.07 & 2.05 & 17.77 & 3.76 & 0.34 & 9.05  & 11.28 & 4.78 \\ 
grep-2.9       & 0    & 6.19 & 0.47 & 1.90  & 4.76  & 0.47 & 3.80  & 1.90  & 1.90 & 7.61  & 2.38 & 0    & 8.57  & 2.38 & 0    & 10.47 & 5.23  & 0.47 \\
gzip-1.4       & 0.58 & 7.01 & 1.75 & 1.75  & 12.86 & 1.16 & 3.50  & 8.18  & 1.16 & 15.78 & 2.92 & 1.16 & 17.54 & 1.75 & 0    & 17.54 & 15.78 & 1.16 \\
lapack-3.3.1   & 2.60 & 5.77 & 0.40 & 3.11  & 5.06  & 1.03 & 4.66  & 3.47  & 1.62 & 7.55  & 1.06 & 0    & 9.24  & 1.06 & 0.81 & 16.11 & 7.09  & 1.34 \\
make-3.82      & 2.61 & 0.52 & 0    & 1.82  & 6.26  & 1.82 & 1.56  & 8.09  & 1.82 & 11.74 & 0.52 & 0    & 6.52  & 2.34 & 1.56 & 12.27 & 4.43  & 0.78 \\
tar-1.26       & 4.53 & 3.27 & 0    & 5.28  & 2.77  & 0    & 2.77  & 2.01  & 0.75 & 7.05  & 0.50 & 0    & 7.05  & 0.25 & 0    & 9.82  & 7.05  & 1.51 \\

%% file: table_time.tex
a2ps-4.14        & 23 & 74 & 34 & 115 & 162 \\    
gawk-4.0.0       & 15 & 46 & 12 & 40 & 50 \\      
gnuchess-6.0.0   & 50 & 220 & 81 & 312 & 351 \\   
gnugo-3.8        & 77 & 159 & 92 & 766 & 1493 \\  
grep-2.9         & 41 & 85 & 22 & 65 & 122 \\     
gzip-1.4         & 22 & 268 & 91 & 303 & 230 \\   
lapack-3.3.1     & 294 & 3740 & 3773 & 8159 & 10351 \\
make-3.82        & 67 & 108 & 53 & 109 & 257 \\   
tar-1.26         & 37 & 218 & 115 & 253 & 396 \\  

%% file: table_domain.tex
a2ps-4.14 & 12.74 & .78 & 0 & 21.64 & 0 & 2.13 & 18.94 & 0 & .93 & 90.47 & 0 & 0 & 0 & .72 & .36 & .77 & 0 & 0 \\ \hline
gawk-4.0.0 & 21.34 & 0 & 0 & 26.96 & 0 & 0 & 17.97 & 0 & 0 & 88.76 & 0 & 0 & 0 & 4.44 & 0 & 0 & 0 & 0 \\ \hline
gnuchess-6.0.0 & 5.99 & 5.78 & 2.47 & 12.67 & 3.68 & 2.24 & 14.87 & 0 & 0 & 83.43 & 0 & 0 & 0 & 2.23 & 0 & .20 & 3.47 & 0 \\ \hline
gnugo-3.8 & 18.75 & 2.08 & 2.08 & 22.50 & 1.66 & 1.11 & 10.86 & 0 & 1.12 & 71.27 & .21 & 1.29 & 0 & .47 & 0 & 0 & 3.69 & 0 \\ \hline
grep-2.9 & 3.30 & 0 & 0 & 8.26 & 0 & 0 & 8.26 & 0 & 0 & 61.74 & 0 & 0 & 0 & .44 & 0 & 0 & 0 & 0 \\ \hline
gzip-1.4 & 21.16 & 2.18 & 0 & 32.84 & .72 & 1.45 & 26.27 & 0 & 0 & 80.29 & 0 & 0 & 0 & 0 & 0 & 0 & 8.75 & 0 \\ \hline
lapack-3.3.1 & 11.84 & 5.67 & .85 & 78.96 & 2.16 & 2.99 & 85.03 & 0 & 0 & 94.46 & 0 & .09 & .09 & 3.22 & .47 & 0 & 4.25 & 0 \\ \hline
make-3.82 & 6.50 & 4.00 & 5.50 & 6.52 & 4.34 & 5.97 & 11.94 & 0 & 0 & 46.50 & 0 & 0 & 0 & 2.29 & 0 & 0 & 2.98 & 5.47 \\ \hline
tar-1.26 & 5.17 & 4.20 & 0 & 9.70 & 3.23 & .97 & 9.38 & 0 & 0 & 62.13 & 0 & 0 & 0 & 3.31 & 0 & 0 & 4.91 & 0 \\ \hline